\documentclass{article}

\usepackage[english]{babel}
\usepackage{ifthen}
\usepackage[letterpaper,top=3cm,bottom=3cm,left=4cm,right=4cm,marginparwidth=3cm]{geometry}

\usepackage{amsmath,amssymb,amsthm}
\usepackage{graphicx}
\usepackage[colorlinks=true, allcolors=blue]{hyperref}

\usepackage{graphicx}
\usepackage[colorinlistoftodos]{todonotes}

\def\Exp{{\mathbb{E}}}
\newcommand{\E}[1]{\mbox{$\Exp[\,#1\,]$}}

\def\Var{{\mathbb{V}\rm ar}}

\newcommand{\bigO}[1]{\mbox{$\textup{O}(\,#1\,)$}}

\def\nump{\textup{\#P}}

\usepackage{thmtools}
\usepackage{thm-restate}
\usepackage{pgfgantt}
\newtheorem{theorem}{Theorem}
\newtheorem{definition}{Definition}
\newtheorem{lemma}{Lemma}
\newtheorem{fact}{Fact}

\newcommand{\red}[1]{\textcolor{black}{#1}}

\title{Minimizing Completion Times of Stochastic Jobs on Parallel Machines is Hard}
\author{Benjamin Moseley$^1$, Kirk Pruhs$^2$, Marc Uetz$^3$, Rudy Zhou$^4$\\[2ex]
{\small ${}^1$Carnegie Mellon University, Pittsburgh, PA, USA}\\ 
{\small ${}^2$University of Pittsburgh, Pittsburgh, PA, USA}\\ 
{\small ${}^3$University of Twente, Enschede, NL}\\
{\small ${}^4$Microsoft, Washington DC, USA}}

\begin{document}
\maketitle

\begin{abstract}
This paper considers the scheduling of stochastic jobs on parallel identical machines to minimize the expected total weighted completion time. While this is a classical problem with a significant body of research on approximation algorithms over the past two decades, constant-factor performance guarantees are currently known only under very restrictive assumptions on the input distributions, even when all job weights are identical. This algorithmic difficulty is striking given the lack of corresponding complexity results: to date, it is conceivable that the problem could be solved optimally in polynomial time. We address this gap with  hardness results that demonstrate the problem's inherent intractability. For the special case of discrete two-point processing time distributions and unit weights, we prove that deciding whether there exists a scheduling policy with expected cost at most a given threshold is \nump-hard. Furthermore, we show that evaluating the expected objective value of the standard (W)SEPT list scheduling policy is itself \nump-hard. These are the first hardness results for scheduling independent stochastic jobs and min-sum objective that do not merely rely on the intractability of the underlying deterministic counterparts.
\end{abstract}

\section{Introduction}

This paper addresses a long-standing open question in stochastic scheduling theory: whether the computational complexity of stochastic parallel machine scheduling fundamentally differs from that of its deterministic counterpart, even in the most basic settings. In deterministic scheduling on parallel identical machines, minimizing the total completion time $\sum_j C_j$ when all jobs have equal weight is solvable in polynomial time by scheduling jobs in shortest processing time (SPT) order~\cite{BCS74}. Here $C_j$ is the completion time of job $j$ under a scheduling policy. By contrast, for the stochastic version of the same problem, where processing times are random and revealed only upon completion, the computational complexity of computing an optimal scheduling policy has remained unresolved since the early 1980s.

Despite decades of research on stochastic scheduling,  
computational hardness results were previously known only for more structured problems as discussed in more detail below.
This paper provides the first evidence that the classical stochastic parallel machine scheduling problem for minimizing $\mathbb{E}[\sum_j C_j$] is computationally intractable, 
thereby separating the stochastic problem from its deterministic counterpart.

We consider a standard stochastic scheduling model with $m$ \emph{parallel identical machines} and $n$ \emph{non-preemptive} jobs. Each job $j$ must be assigned to a single machine and, once started, runs to completion without interruption. The processing time $x_j$ of job $j$ is unknown in advance and is modeled as a nonnegative random variable $X_j$, independently distributed across jobs. Following recent work on approximation algorithms for stochastic scheduling, we focus on a discrete variant of the problem in which each $X_j$ has finite support, and in particular we assume that each distribution has \emph{support size at most two}. Each job $j$ additionally may have a non-negative weight $w_j\ge 0$.

Under a scheduling policy, let $S_j$ and $C_j$ denote the (random) start and completion times of job $j$, respectively. By definition,
$\mathbb{E}[C_j] = \mathbb{E}[S_j] + \mathbb{E}[X_j]$.
The objective is to find, or evaluate, a scheduling policy that minimizes the expected weighted sum of completion times,
\[
\mathbb{E}\!\left[\sum\nolimits_j w_j C_j\right].
\]

Optimal scheduling policies for this problem are known only in highly restricted special cases. When all weights are equal and processing times follow exponential distributions, the greedy list scheduling policy SEPT is optimal~\cite{WP80,BDF81}. \red{A list scheduling policy pre-computes an ordering of the jobs, and whenever a machine falls idle, it schedules the next job in this order. For SEPT, this order is Shortest Expected Processing Time first.}
More generally, SEPT is still an optimal policy if the processing time distributions are pairwise stochastically comparable~\cite{WVW86}.
With exponentially distributed jobs, the optimality of SEPT also extends to weighted jobs whenever the SEPT order is agreeable with a non-decreasing order of job weights~\cite{K87}.
Outside of such special settings, however, it is well understood that optimal policies must be \emph{adaptive}, reacting dynamically to realized processing times~\cite{Uetz03,SSU2016,EFMM2019}. Moreover, the adaptivity gap between non-adaptive and adaptive policies can grow proportionally to the coefficient of variation of the processing time distributions~\cite{SSU2016,EFMM2019}.

Beginning with~\cite{MSU99}, a substantial literature has developed in which various approximation algorithms have been proposed for stochastic scheduling. Many of these are based on linear programming relaxations of the optimal adaptive policy, either in designing the algorithm itself, or in its analysis~\cite{MSU99,SU2005,MUV2006,SSU2016,BBC2016,GMUX2020,GMUX2021,J2023}. These approximation guarantees scale with the largest squared coefficient of variation of the random variables $X_j$, defined as $\Var[\,X_j\,]/\E{X_j}^2$. For the special case of unit weights, an exception is the work of~\cite{IMP2015}, whose performance guarantee instead depends on the number of machines $m$. More recently, for two-point Bernoulli processing times, sublinear-in-$m$ approximation guarantees have been obtained that do not depend on second moments~\cite{GMZ2023,AHSU2025}: However, the approximation guarantee of~\cite{GMZ2023}, while sublinear in $m$, contains large poly-logarithmic factors in the number of jobs~$n$, and the guarantee of~\cite{AHSU2025} is \bigO{\log n}, but only quasi-polynomial time.

Despite the sophistication of these approximation results, strong lower bounds on the computational complexity of stochastic scheduling have remained largely absent. When job weights are arbitrary, the problem is NP-hard, since it generalizes deterministic scheduling on parallel machines, which is NP-hard~\cite{GLLR-K1979}. However, for the unweighted case where $w_j = 1$ for all jobs $j$, deterministic scheduling is optimally solvable by SPT~\cite{BCS74}. Quite strikingly, the computational complexity of the corresponding stochastic problem ---minimizing $\mathbb{E}[\sum_j C_j]$ on parallel identical machines--- has remained open for over four decades. {In particular, it is still open if the stochastic problem can be solved optimally in polynomial time.}
In summary, the lurking fundamental question is:
\begin{quote}
\emph{Is it computationally hard to compute an optimal scheduling policy for minimizing expected total completion time on parallel identical machines?}
\end{quote}
Despite sustained interest in stochastic scheduling since the 1980s, this question has not been resolved. This paper gives a partial answer, testifying the problem being computationally harder than the corresponding deterministic counterpart.

\subsection{Our Results}

This paper provides the first indication that this stochastic scheduling problem is unlikely to admit a polynomial-time solution, even when all job weights are equal. As a main result the paper shows that the decision problem of whether there exists a scheduling policy whose expected cost $\E{\sum_j C_j}$ is at most a given threshold is \nump-hard. 
The reduction is of the Turing type, meaning that if we were able to decide the existence of such a policy in polynomial time,
this could be used as a subroutine to solve any other counting problem in \nump\ in polynomial time.
This hardness result holds even under 
\emph{restricted stochasticity}, namely with independent two-point processing time distributions. 

As a matter of fact, we derive that main result from another, maybe even more surprising result -- namely that already computing the expected cost of the simple list scheduling policies WSEPT (Weighted Shortest Expected Processing Time first) and SEPT (Shortest Expected Processing Time first) is generally \nump-hard. 
Recall that WSEPT, list scheduling in non-increasing order of $w_j/\E{X_j}$, is known to be optimal on a single machine for both deterministic~\cite{Smith56} and stochastic processing times~\cite{Rothkopf66}, 
and WSEPT yields constant-factor approximations when the coefficient of variation is bounded by a constant~\cite{MSU99,JS2018}. However the complexity of computing its expected cost, and also that of SEPT for the special case of unit weights, was open.

Our results are based on increasingly refined versions of the same basic construction, starting with the simplest one for WSEPT, adapting it to SEPT, and finally arguing about an optimal scheduling policy. Each of them yields a reduction of the Turing type. 
Altogether, the results highlight a fundamental barrier to computing optimal scheduling policies, and also to evaluating the performance of even the most basic greedy scheduling policies.

\subsection{Scheduling Polices, Adaptivity and Deliberate Idleness}
We briefly discuss scheduling policies in this section. We view a policy as a Markov decision process, and refer to \cite{MRW84} for more details and other representations of policies, e.g., as specific functions mapping processing times to completion times. The states of the Markov decision process are a ``snapshot'' of the partial schedule at a given time. This snapshot is completely described by all processing time distributions $X_j$, the time $t\ge 0$, the set of jobs started prior to $t$ along with their realized start times $S_j$, plus potentially their realized completion times $C_j$ if they have completed by time $t$, so if $C_j \leq t$. Then the set of yet unscheduled jobs is known at any time, too.

Further, for all jobs $j$ started at time $S_j\le t$ but not yet completed by time $t$, we know the remaining processing time of $j$ is distributed as $X_j| X_j> t-S_j$. Following \cite{MRW84}, the action of a scheduling policy, given a state at time $t$, is a subset of jobs to be scheduled at $t$ on a set of idle machines, together with a tentative next decision time $t^{\text{next}} > t$. The transition to the next state is then determined by proceeding to time $t^{\text{next}}$, or the next job completion time -- whichever is smaller. The policy must be non-anticipatory, meaning that its scheduling decisions can only be based on the current state at any time $t$. Note that even a greedy list scheduling policy like (W)SEPT is adaptive in the sense that the job-machine assignments generally depend on realized processing times. Another level of adaptivity is when a policy does not even follow a fixed priority list. As discussed, it is well known that fixed priority lists are generally not sufficient for optimality. 

We point out one peculiarity which is relevant to our constructions in this paper, which is when jobs have realized processing time zero. For this, it is more convenient and cleaner to relax the definition of the next decision time to $t^{\text{next}}\ge t$. Then a scheduling policy could start a job $j$ at time $t$ on some idle machine, realize that its processing time equals zero, and the next action could schedule yet another job $k$ at the same time $t$ and on the same machine.

Another important detail is that in general, a scheduling policy could decide to leave machines deliberately idle. Thus, even though there are unscheduled jobs and at least one idle machine, it decides to not schedule any job, and instead proceed to time $t^{\text{next}} > t$ or the next job completion time. It is known that there exist instances where the optimal policy must idle~\cite{Uetz03}. We finish this section with one simple observation for finite discrete processing time distributions that will become relevant in one of the proofs. The observation is mentioned also in \cite{MRW84}. We refer to Appendix~\ref{app:fact1} for a formal proof. 
\begin{fact}\label{fact:minimal_idleness}
    Suppose all processing time distributions are finite discrete. Then whenever a scheduling policy $\Pi$ leaves a machine deliberately idle at some time $t$, one may assume w.l.o.g.\ that the length of that idle period is at least as long as the time between $t$ and the next possible job completion.
\end{fact}
Here, make sure to distinguish between the next \emph{possible} job completion and the next job completion.

\subsection{Relation to Existing Lower Bounds}
Hardness results for stochastic scheduling problems are known only in more structured settings, for non-adaptive scheduling policies, and moreover for the makespan objective which is arguably more amenable to lower bound constructions. We briefly discuss these earlier lower bounds and how our contribution differs from them.

First, there is a PSPACE-hardness result for another variant of a stochastic scheduling problem~\cite[Thm.~19.7]{PapadComplexityBook1994}. This is for stochastic scheduling with so-called alternate precedence constraints and makespan objective for a given subset of mandatory jobs. Alternate precedence constraints means that we are given a directed acyclic graph, and a job can only be started if any one of its predecessors has been finished. Non-mandatory jobs only need to be processed if needed as predecessors of mandatory jobs. This rich combinatorial structure allows to build a reduction from the PSPACE-hard stochastic satisfiability problem.

Second, in precedence-constrained stochastic project scheduling, a.k.a.\ PERT (Program Evaluation and Review Technique), computing the expected makespan is \nump-hard~\cite{Hagstrom88}. Here,  the \nump-hardness follows by reducing from reliability problems in directed graphs~\cite{Valiant1979}, and  precedence constraints are crucial to do that.

Third, a computationally hard problem with less combinatorial structure is the problem to compute the probability for the sum of independent Bernoulli trials being bounded by some threshold.  This problem can be equivalently interpreted as a single machine stochastic scheduling problem with makespan objective, and it is \nump-hard, too~\cite{KRT00}. This latter result has been used to show that also the computation of the expected makespan of stochastic jobs on parallel machines is \nump-hard, however under the additional assumption that the jobs are pre-assigned to the machines~\cite{GKNS20}. This latter assumption makes the corresponding policy non-adaptive. Non-adaptive policies are often simpler to analyze and more amenable to gadget-style reductions, since they correspond to selecting a fixed subset of jobs for each machine. In contrast, an adaptive policy corresponds to a stochastic process which constructs a job-to-machine assignment which generally depends on realized processing times.

In summary, prior hardness results are known only for the makespan objective, and further they require additional combinatorial constraints (precedence constraints) or are restricted to weaker, non-adaptive policies. To the best of our knowledge, our work is the first hardness result for the min-sum objective and the class of adaptive scheduling policies. Showing hardness for adaptive policies is more challenging due to the absence of any additional combinatorial structure.
However it is particularly relevant for the problem at hand, since it is well known that the adaptivity gap can be large for stochastic parallel machine scheduling \cite{SSU2016}, and the state-of-the-art algorithms for this problem compute adaptive policies \cite{GMZ2023,AHSU2025}. A challenge in the proposed proof is therefore to construct input instances that allow to infer certain structural properties of the optimal \emph{adaptive} policies.

\subsection{Paper Organization}
We proceed by first giving the hardness proof for computing $\E{\sum_j w_jC_j}$ for WSEPT. Due to the presence of arbitrary weights, this is the simplest construction. We then proceed to adjust the same construction, first to show hardness of  computing $\E{\sum_j C_j}$ for SEPT, and finally to show hardness of computing $\E{\sum_j C_j}$ for an optimal policy. The latter will follow by an additional, small adjustment in the construction, that allows to prove that SEPT must be an optimal policy.

\section{Hardness of evaluating WSEPT and SEPT}

For the rest of the paper, for simplicity, we assume the objective is to minimize the (weighted) sum of expected starting times $\E{\sum_j w_jS_j}$ instead of  $\E{\sum_j w_jC_j}$, because   $\E{\sum_j w_jC_j}$ $=$ $\E{\sum_j w_jS_j} + \E{\sum_j w_jX_j}$. This is w.l.o.g.\ as long as we talk about exact computation and not approximation, because the jobs are non-preemptive. Therefore,  $\E{C_j}=\E{S_j}+\E{X_j}$, and the latter term is a constant independent of the scheduling policy.

The reductions that are to follow use the \nump-hardness of the problem to count the number of feasible solutions for a knapsack constraint. This problem is one of the most basic \nump-hard counting problems~\cite{Simon77}, see also~\cite{Jerrum2003} and~\cite{GopalanEtAl2011}.
\begin{definition}[{\sc Knapsack}]
An instance of {\sc Knapsack} has $n\ge 2$ items with integer sizes $s_i$ and an integer knapsack bound $B$, where  $1\le s_i\le B$ and $B\ge 2$. A feasible solution is a subset of items $W\subseteq[n]$ that fulfills the knapsack constraint $\red{s(W):=}\sum_{i\in W} s_i\le B$.
\end{definition}

We will use the following lemma where we consider restricted instances on the {\sc Knapsack} counting problem. Its proof can be found in Appendix~\ref{app:Lemma1}.
\begin{lemma}\label{lem:pack}
    The problem to count the number of feasible solutions for any instance $(s_1,\dots,s_n, B)$ of problem {\sc Knapsack} is a \nump-hard problem,  and this is true also if we assume that $B+1 <\sum_{j\in[n]} s_j\le 3B/2$.
\end{lemma}
\red{The reason to include the $3B/2$ bound on the total size of all items in Lemma~\ref{lem:pack} is a volume argument that we later use in the proof of Lemma~\ref{lem:Delta_SEPT}. In fact instead of $3B/2$, any other constant in between $B$ and $2B$ would work as well.} 

\subsection{Hardness of WSEPT}

The goal of this section is to prove the following theorem. 
\begin{theorem}\label{thm:WSEPT} Consider parallel machine stochastic scheduling to minimize expected cost $\E{\sum_j w_jC_j}$. When using WSEPT, \red{list scheduling in non-increasing order of $w_j/\E{X_j}$}, computing $\E{\sum_j w_j C_j}$ is \nump-hard.
\end{theorem}

For the reduction, take any instance $(s_1,\dots,s_n, B)$ of {\sc Knapsack} with items sizes $s_j$ and knapsack capacity $B$. 

\medskip \noindent 
\textbf{Defining Knapsack Jobs:} First we define $ n $ ``knapsack jobs'' with stochastic processing times $X_j$ which are either $ s_j $ or $1/n $, each with probability $ {1}/{2}$, and independent of each other. Let us refer to a processing time realization $X_j=s_j$ as \emph{long}, and \emph{short} otherwise. This definition is inspired from the proof of Theorem~2.1 of~\cite{KRT00}. We note that instead of choosing $1/n$ for short realizations, one could choose another function of $n$ here; the requirement for our proofs to work is that the sum of $n$ short knapsack jobs should not exceed $1$. 
 Observe that each realization of the vector $(X_1,\dots,X_n)$ of all processing times of knapsack jobs has the same probability $1/2^n$. Therefore, for each
subset $W \subseteq [n] $, with probability $1/2^n$ the jobs $j\in W$ are long and $j\not\in W$ short, in which case the total processing time of all knapsack jobs $X([n])$ lies in the interval $[s(W),s(W)+1]$.

Denote a realization of processing times of the $n$ knapsack jobs a YES realization, whenever the set of jobs $W$ that turn out long corresponds to a feasible solution $W$ for the knapsack problem, i.e., $s(W)=\sum_{j\in W} s_j\le B$.  Observe that this implies $X([n])=\sum_{j\in[n]} X_j < B+1$. Denote a realization of processing times of the $n$ knapsack jobs a NO realization otherwise. 

\medskip

\begin{figure}[t]
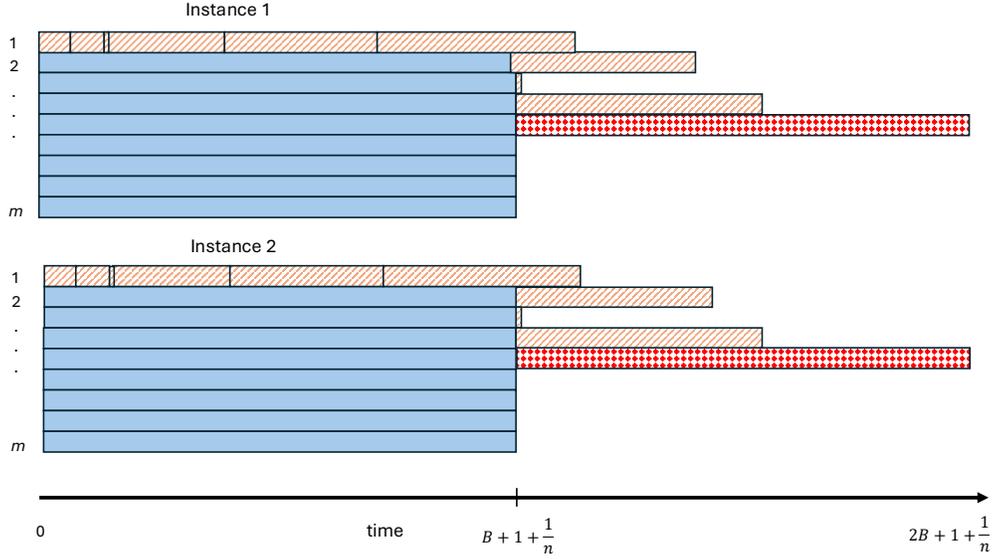

    \centering
\begin{ganttchart}[
expand chart=0.9\textwidth,
    y unit chart=0.3cm,
    include title in canvas=false,
    hgrid,
    /pgfgantt/bar label font = \scriptsize,
    y unit title=.25cm,
    title/.style={draw=none, fill=none},
    bar/.style={line width=0.4pt, draw=black} 
]{0}{200}  

\gantttitle{Instance 1}{200}\\
  \ganttbar[bar/.append style={fill=red!40}]{1}{0}{115}
  \ganttbar[inline,bar/.append style={fill=red!40}]{}{0}{78} 
  \ganttbar[inline,bar/.append style={fill=red!40}]{}{0}{46} 
  \ganttbar[inline,bar/.append style={fill=red!40}]{}{0}{28}
  \ganttbar[inline,bar/.append style={fill=red!60}]{}{0}{16}
  \ganttbar[inline,bar/.append style={fill=red!40}]{}{0}{15}
  \ganttbar[inline,bar/.append style={fill=red!40}]{}{0}{5}
	\\
  \ganttbar[bar/.append style={fill=red!40}]{2}{0}{145} 
  \ganttbar[inline,bar/.append style={fill=blue!40}]{}{0}{98} \\
  \ganttbar[bar/.append style={fill=red!40}]{}{0}{150} 
  \ganttbar[bar/.append style={fill=blue!40}]{}{0}{100} \\
  \ganttbar[bar/.append style={fill=gray!20}]{}{0}{200} 
  \ganttbar[inline,bar/.append style={fill=blue!40}]{}{0}{100} \\
  \ganttbar[bar/.append style={fill=blue!40}]{}{0}{100} \\
  \ganttbar[bar/.append style={fill=blue!40}]{$\vdots$}{0}{100} \\
  \ganttbar[bar/.append style={fill=blue!40}]{}{0}{100} \\
  \ganttbar[bar/.append style={fill=blue!40}]{}{0}{100} \\
  \ganttbar[bar/.append style={fill=blue!40}]{$m$}{0}{100} 

   \ganttvrule[vrule/.style={black!70, dashed,  ultra thick}]{}{100}
   \end{ganttchart}
   \begin{ganttchart}[
expand chart=0.9\textwidth,
    y unit chart=0.3cm,
    /pgfgantt/bar label font = \scriptsize,
    include title in canvas=false,
    hgrid,
    y unit title=.25cm,
    title/.style={draw=none, fill=none},
    bar/.style={line width=0.4pt, draw=black} 
]{0}{200}  

\gantttitle{Instance 2}{200}\\
   \ganttbar[bar/.append style={fill=red!40}]{1}{0}{115}
  \ganttbar[inline,bar/.append style={fill=red!40}]{}{0}{78} 
  \ganttbar[inline,bar/.append style={fill=red!40}]{}{0}{46} 
  \ganttbar[inline,bar/.append style={fill=red!40}]{}{0}{28}
  \ganttbar[inline,bar/.append style={fill=red!60}]{}{0}{16}
  \ganttbar[inline,bar/.append style={fill=red!40}]{}{0}{15}
  \ganttbar[inline,bar/.append style={fill=red!40}]{}{0}{5}
	\\
  \ganttbar[bar/.append style={fill=red!40}]{2}{0}{147} 
  \ganttbar[inline,bar/.append style={fill=blue!40}]{}{0}{100} \\
  \ganttbar[bar/.append style={fill=red!40}]{}{0}{150} 
  \ganttbar[bar/.append style={fill=blue!40}]{}{0}{100} \\
  \ganttbar[bar/.append style={fill=gray!20}]{}{0}{200} 
  \ganttbar[inline,bar/.append style={fill=blue!40}]{}{0}{100} \\
  \ganttbar[bar/.append style={fill=blue!40}]{}{0}{100} \\
  \ganttbar[bar/.append style={fill=blue!40}]{$\vdots$}{0}{100} \\
  \ganttbar[bar/.append style={fill=blue!40}]{}{0}{100} \\
  \ganttbar[bar/.append style={fill=blue!40}]{}{0}{100} \\
  \ganttbar[bar/.append style={fill=blue!40}]{$m$}{0}{100} 

   \ganttvrule[vrule/.style={black!70, dashed,  ultra thick}]{$B+1+\frac{1}{n}$}{100}
   \end{ganttchart}
    \caption{Illustration of an instance with $n=m=9$: Schedules for the two scheduling instances with blocker jobs (blue), knapsack jobs (red), and dummy job (light gray) for some NO realization of the processing times. Note that except for the knapsack job on machine no.~2, all start times are identical.}
    \label{fig:instances}
\end{figure}

Inspired by the proof strategy of \cite{GKNS20}, we define two slightly different instances of the stochastic scheduling problem; see Figure~\ref{fig:instances} for an illustration of sample schedules produced by WSEPT on these two instances.

\medskip \noindent 
{\bf Instance 1}: There are the above described $ n $ knapsack jobs, each with weight $w_i =  1$. There are $ m $ machines, where $m= n$, and $ m-1 =n-1 $ ``blocker jobs''. One of the blocker jobs has deterministic processing time $X_j=B+1$, while all remaining $m-2$ blocker jobs have deterministic processing time $X_j=B+1+1/n$. Choose the weight of all blocker jobs as $w_j=6B$.  Finally, there is one deterministic dummy job with weight $ w_j=1 $ and deterministic processing time $X_j=B$. 

\medskip
 \noindent {\bf Instance 2}: This instance is identical to Instance~1, with the only difference that \emph{all} $ m-1 $ blocker jobs have deterministic processing time $ B+1+1/n $.

\medskip

Now we analyze these two instances. 

\begin{fact}
    For both instances, WSEPT schedules first all blocker jobs, then all knapsack jobs, and finally the dummy job.
\end{fact}
\begin{proof}
     Observe that all blocker jobs have $w_j/\E{X_j}\ge 6B/(B+1+1/n)>2$, and for all knapsack jobs we have $w_j/\E{X_j}\le 2$. Moreover, since $w_j/\E{X_j}> 1/B$ for the knapsack jobs, 
     WSEPT has to schedule the dummy job at last.
\end{proof}
\medskip

We next introduce a bit of notation in order to be able to talk about the difference in expected costs under WSEPT for Instances~1 and~2. First, denote by $ E_1 $ and $ E_2 $ the total expected costs $\E{\sum_j w_j S_j}$ under WSEPT for Instances 1 and 2, respectively. Further, 
for $W\subseteq [n]$, denote by $S^{\red{1}}_j(W)$ and $S^{\red{2}}_j(W)$ the start time of a job $j$ under WSEPT for Instances~1 and~2, respectively, assuming that exactly the knapsack jobs  $j\in W$ turn out long.  By definition, we then have
\[
E_i= \frac{1}{2^n}\sum\nolimits_{W\subseteq[n]} \sum\nolimits_j w_jS_j^{\red{i}}(W), \quad i=1,2\,,
\]
because each realization $W$ has equal probability $1/2^n$.
Next, for each realization $W$, define the difference of the costs of the two schedules, 
\[
\Delta(W):= \sum\nolimits_j w_j\left(S_j^{\red{2}}(W)-S_j^{\red{1}}(W)\right)\,.
\]
Then, the difference in expected costs under WSEPT for Instances~1 and~2 is 
\[
\Delta:=E_2-E_1 = \frac{1}{2^n}\sum\nolimits_{W\subseteq [n]}\Delta(W)\,.
\]

\begin{lemma}
    For each YES realization $W$ of the processing times of the knapsack jobs, $\Delta(W)=0$, hence there is no contribution to the difference $\Delta$.
\end{lemma}
 \begin{proof}
     All blocker jobs start at time 0.  Therefore all but one machine is blocked until time at least $B+1$ with blocker jobs. This is true for both instances. Hence for any YES realization $W$ of the knapsack jobs, these jobs start in the same WSEPT order on the only available machine from time $0$ on, in both instances. Since for a YES realization, $X([n])< B+1$, also the dummy job starts on that machine. Hence for each YES realization $W$, the start time of every single job is identical in both instances. Therefore, $\Delta(W)=0$ as claimed.
 \end{proof}

\begin{lemma}\label{lem:NO-WSEPT}
    For each NO realization $W$ of the processing times of the knapsack jobs, $\Delta(W)=1/n$, hence  the contribution to the difference $\Delta$ is $1/(n2^n)$.
\end{lemma}

 \begin{proof}
    Consider any NO realization $W$ of the knapsack jobs. Again all blocker jobs are started at time 0, and let us refer to the one machine that is free from blocker jobs as the \emph{free machine}. (Machine no.\ 1 in Figure~\ref{fig:instances}.) 
    First note that, again, in both instances the same set of jobs will be scheduled from time 0 on in the same WSEPT order on the free machine before time $B+1$. Consider the prefix $Pr\subseteq [n]$ of knapsack jobs started on the free machine and with start time at most $B+1$. Note that this set of jobs, and their start times, are identical in both instances. Since we consider a NO realization $W$, we have $s(W)\ge B+1$, hence the total processing volume of the prefix jobs is $X([Pr])\ge B+1$.
    
    Consider first the case that $Pr=[n]$, meaning that all knapsack jobs are started on the free machine and only the dummy job is left to schedule. Since we assumed that $s([n])>B+1$, by definition of prefix $Pr$ we must have that 
    \begin{equation}\label{eq:prefix_lb}X([Pr])\ge B+1+1/n\,.\footnote{Note that \eqref{eq:prefix_lb} might hold with equality, namely when the set of long knapsack jobs equals  $W=[n]\setminus j$ with $s(W)=B+1$, and $X_j=1/n$.} \end{equation}
    Consequently, in Instance 1 the dummy job will be scheduled at time $B+1$, which is the earliest time when a machine falls idle. In Instance 2, however, the dummy job can be started at time $B+1+1/n$ only. Hence, only the dummy job has different start times and $\Delta(W)=1/n$, as claimed.
    
    Next consider the case where $Pr\subset [n]$, meaning that at time $B+1$ there are some knapsack jobs left to be scheduled, plus the dummy job. By definition of prefix $Pr$ we again have that \eqref{eq:prefix_lb} holds. Then in Instance 1, the first of the jobs $[n]\setminus Pr$ will be started at the earliest time when a machine is available, which is at time $B+1$. As its processing time is at least $1/n$, the next earliest start time on any machine is $B+1+1/n$, but at that time all the remaining machines are available. Consequently,   
    all remaining jobs including the dummy job, which is at most $n-1$ jobs because $|Pr|\ge 2$,  are started at time $B+1+1/n$.  In Instance 2, in contrast, \emph{all} jobs $[n]\setminus Pr$ and the dummy job are started at time $B+1+1/n$. That means that, per NO realization $W$, it is precisely one job that will be started at time $B+1$ in Instance 1, and at time $B+1+1/n$ in Instance 2. All other jobs $[n]\setminus Pr$ share the same start time $B+1+1/n$, in both instances.  Hence for each NO realization $W$, again $\Delta(W)=1/n$, as claimed. 
\end{proof}

\begin{proof}[Proof of Theorem~\ref{thm:WSEPT}]
To finish the proof of Theorem~\ref{thm:WSEPT}, 
assume that we could compute $\E{\sum_j w_j S_j} $ when using WSEPT, in time polynomial in the input size of the problem.  Define $ k $ as the number of sets $W$ that do not fit into the knapsack for the given knapsack instance. By the two previous lemmas, we get $\Delta=k/(n2^n)$, or in other words, $k=n2^n\Delta$.
Hence the number of feasible solutions for the knapsack constraint equals $2^n(1-n\Delta)$ and since evaluating WSEPT allows to compute $\Delta=E_2-E_1$, the proof is finished.
\end{proof}

\subsection{Hardness of SEPT}

In this section, we extend the ideas of the prior section to the unweighted case. With a little more care in defining the set of blocker jobs, we next show that the same result even holds when all weights $w_j$ are equal to~1. 
\begin{theorem}\label{thm:SEPT}
    Consider  parallel machine stochastic scheduling  to minimize expected cost $\E{\sum_j C_j}$. When using SEPT, \red{ list scheduling in non-decreasing order of $\E{X_j}$}, computing $\E{\sum_j  C_j}$ is \nump-hard.
\end{theorem}

As in the proof of Theorem~\ref{thm:WSEPT}, for an instance $(s_1,\dots, s_n,B)$ of the knapsack problem, we define $n$ stochastic knapsack jobs with processing times $X_j$ equal   $s_j$ or $1/n$, each with probability $1/2$, independent of each other. Again we introduce two instances with $ m-1 $ blocker jobs, the main difference being that the blocker jobs will now be independent jobs that follow a Bernoulli distribution.

\medskip

\noindent {\bf Instance 1}: Next to the  $n$ knapsack jobs as in the proof of Theorem~\ref{thm:WSEPT}, there is a single  stochastic blocker job with possible processing times $X_j$ equal to $ 0 $ or $ B +1 $. All other $ m-2 $ Bernoulli blocker jobs have possible processing times $X_j$ equal to $ 0 $ or $ B+1 +1/n $. Each blocker job has processing time $0$ with probability $ 1- q $, and processsing time $B+1$, resp.\ $B+1+1/n$, with probability $ q $, while $0<q<1$. All blocker jobs are independent.  Finally, there is again a  dummy job with deterministic processing time $X_j=B+1$.

\medskip

\noindent {\bf Instance 2}: Is identical to Instance~1, with the only difference that \emph{all} $ m-1 $ blocker jobs have processing times $X_j$ either $0$ or $ B+1 +1/n$. The independent probabilities for being short or long are the same as in Instance~1.

\medskip
Note that the expected processing time of any blocker job is at most $q(B+1+1/n)<1/2$  whenever we choose $q<1/(2(B+1+1/n))$. Also note that the encoding length of $q$ is \bigO{\log B}.  
\begin{fact}
    By choosing $q$ small enough, e.g., $q= 1/(\red{4}B)$, SEPT must first schedule all blocker jobs, in both instances. Moreover the size of the stochastic scheduling instances is polynomial in the size of the instance of {\sc Knapsack}.
\end{fact}

To be able to use the same proof strategy as before, we basically argue that we may assume  w.l.o.g.\ that all blocker jobs turn out long. This is made precise in the following lemma, which formalizes that all realizations where \emph{not} all blocker jobs are long, don't matter. For this to work out, we rely on the fact that by Lemma~\ref{lem:pack}, we may assume that $s([n]) \leq 3B/2$.
\begin{lemma}\label{lem:Delta_SEPT}
    Denote by $ E_1 $ and $ E_2 $ the expected costs $\E{\sum_j S_j}$ using SEPT for Instances~1 and~2, respectively, and again define the difference in expected costs as $\Delta:=E_2-E_1\,.$
    Denote by $E_1'$ and $E_2'$ the expected costs $\E{\sum_j S_j}$ using SEPT for Instances~1 and~2, but under the assumption that all blocker jobs are long. Then we have
    \[
    \Delta = q^{m-1}(E_2'-E_1')\,.
    \]
\end{lemma}
\begin{proof}
    In both instances, the probability that all blocker jobs are long equals $q^{m-1}$. The lemma follows because in all other realizations of the lengths of the blocker jobs, and for all possible realizations of the knapsack jobs, the costs of the two SEPT schedules is identical for both instances. To see this, observe that if at least $h\ge 1$  of the blocker jobs turns out short, there are $ h +1 \geq 2 $ machines free to process the knapsack jobs. This happens in either of the two instances with the same probability (it equals ${m-1\choose{h}}q^{m-h-1}(1-q)^{h}$, which is the probability that $h$ of the $m-1$ blocker jobs are short). \red{Now recall that  Lemma~\ref{lem:pack} allowed us to assume $\sum_{j\in[n]} s_j \le {3B}/{2} $. That implies that, in any such case, the total processing volume of the knapsack jobs is at most $3B/2$. But there are at least $ h +1\ge 2$ available machines, a total processing capacity of $(h+1)B\ge 2B$ up to time $B$.
    Hence when using SEPT, no matter 
    what the realizations of knapsack jobs are, the SEPT schedules of the knapsack jobs are identical in both instances. Moreover, since $3B/2<2B\le (h+1)B$, at least one machine will be finished processing knapsack jobs before time $B$. Therefore, the start time of the dummy job must be identical in both instances, too.}
    Abusing notation a bit,  we thus have $\Delta(W)=0$  for all realizations where at least one blocker job is short, and that holds for all realizations $W$ of knapsack jobs. Therefore $\Delta= (1-q^{m-1})\cdot 0 + q^{m-1}\cdot(E_2'-E_1')$.
\end{proof}
That means that we can effectively condition on the case where all blocker jobs turn out long. 
That being said, we get the same lemmas as before, so we keep their formulation informal and do not repeat their proofs here.
\begin{lemma}
    A realization of processing times where all blocker jobs are long, together with a YES realization $W$ of the processing times of the knapsack jobs, contributes nothing to the difference $\Delta$.
\end{lemma}

\begin{lemma}
    A realization of processing times where all blocker jobs are long, together with a NO realization $W$ of the processing times of the knapsack jobs, contributes precisely $q^{m-1}/(n2^n)$ to the difference $\Delta$.
\end{lemma}

\begin{proof}[Proof of Theorem~\ref{thm:SEPT}.]
\red{By the previous lemmas, the only nonzero contribution to the difference $\Delta$ is when all blocker jobs are long, and when the knapsack instance is NO.} So assume that we could compute $\E{\sum_j S_j} $ when using SEPT, in time polynomial in the input size of the problem. 
Using the above three lemmas,  we see that the difference in expected costs for the two instances is $ \Delta = q^{m-1} \cdot \frac{k}{n2^n} $, where $k$ is the number of NO instances for the knapsack problem. Hence using Lemma~\ref{lem:Delta_SEPT}, based on the assumption that we can compute $E_2-E_1$, and observing that the encoding length of $q^{m-1}$, $|q^{m-1}|\in\bigO{m\log B}$, we can compute the quantity $(E_2'-E_1') = (E_2-E_1)/q^{m-1}$ in polynomial time.
Hence the number of feasible solutions of a given instance of the problem {\sc Knapsack}
can be computed as $2^n(1-n(E_2'-E_1'))$. 
\end{proof}

\section{Stochastic scheduling is presumably hard(er)}
    
Here we aim to separate the deterministic scheduling problem to minimize the total completion time on parallel machines from the corresponding stochastic problem. Recall that the deterministic problem can be solved optimally in polynomial time by computing an SPT  schedule~\cite{BCS74}, greedily scheduling by shortest processing time first.  The first step is to show that optimally solving the stochastic problem at least entails a computationally hard (sub) problem in the following sense.
\begin{theorem}\label{thm:OPT}
Consider parallel machine stochastic scheduling to minimize expected cost $\E{\sum_j C_j}$. Then the computation of the minimum expected cost is \nump-hard.
\end{theorem}

\def\alg{\mbox{\sc Alg}}

The proof will rely on the following lemma that specifies how an optimal scheduling policy acts on the instances that were used to prove Theorem~\ref{thm:SEPT}. Effectively, we will show that the optimum policy must follow SEPT,  if we choose the problem parameters more carefully. Here observe that, since not all processing time distributions are stochastically comparable, this observation does not follow from~\cite{WVW86}.

As a first step, we show that the optimum policy must start all blocker jobs at time 0 when we choose the probability $q$, the probability for the blocker jobs being long, small enough.  Intuitively, with small enough $q$ it is better to gamble on \emph{all} blocker jobs being of size $0$, because then all the knapsack jobs can be started at time 0, too. This yields a total cost $0$ for them. Once this is established, and no matter how many of the blocker jobs turn out long, all the remaining jobs are in fact stochastically comparable, and hence it follows from \cite{WVW86} that SEPT is an optimal scheduling policy for all remaining jobs.
Before giving the proof of Theorem~\ref{thm:OPT}, we turn the intuition about the blocker jobs always going first into a formal argument.
\begin{lemma}\label{lem:OPT_equal_ALG}
    Consider Instances~1 or~2 as described in the proof of Theorem~\ref{thm:SEPT}. Assume that $q$, the probability  for long blocker jobs, is chosen small enough, say $q=1/(\red{4}n^2B)$. Then 
    any optimum policy must schedule all blocker jobs at time $0$. 
\end{lemma}

\begin{proof}
Consider a scheduling policy $\Pi$ that does \emph{not} always start all blocker jobs at time~$0$, {and assume it nevertheless minimizes the expected cost}. 
That means that there exists at least one  realization of (some of the) processing times, so that the policy, while still at time $0$, decides to \emph{not} schedule at least one unscheduled  blocker job on an available machine. That means that there is a state and action at time 0 so that $\Pi$ leaves some blocker job $b$ unscheduled, and instead either schedules other, i.e., non-blocker jobs on all remaining available machines, or it deliberately leaves at least one machine idle at time~0.
From here on, we condition on the partial job realizations and sequence of $\Pi$'s actions that has resulted in one such state.
    
We claim that the earliest time when  blocker job $b$ can be started by policy $\Pi$ is at time $1/n$. This is true because, if $\Pi$ schedules non-blocker jobs on all remaining available machines, their processing time is at least $1/n$. On the other hand if $\Pi$ leave at least one machine idle, by
Fact~\ref{fact:minimal_idleness}, we may assume that this idle time is at least $1/n$, too.
We next show how to modify $\Pi$'s action for the selected state, and define an alternative  policy $\Pi'$ that has strictly smaller expected cost. As a consequence, any policy that does not start all blocker jobs at time $0$ for all possible realizations, is suboptimal.

First, observe that the expected remaining cost of  policy $\Pi$, conditioned on the selected state, is at least $1/n + c_\Pi(\text{remain})$, where $c_\Pi(\text{remain})$ denotes the smallest possible expected cost for scheduling all remaining jobs in the given state. Here,  we define the term $c_\Pi(\text{remain})$  so as to \emph{exclude} the existence of  blocker job $b$, which we do because its smallest possible expected cost is already accounted for in the term $1/n$. 
    
Next consider policy $\Pi'$, identical to $\Pi$ except for the action at the  selected state. Instead, $\Pi'$ starts blocker job $b$ at time 0. The idea is that with overwhelming probability, blocker job \red{$b$} has processing time $0$, in which case $\Pi'$ will have strictly smaller expected remaining cost compared to $\Pi$. In the other case this is reverse, but it happens with small enough probability.

To make that formal, we consider two cases: Conditioned on the given state \emph{and} the blocker job $b$ turning out short, $\Pi'$ continues to simulate $\Pi$, with the same expected remaining cost $c_\Pi(\text{remain})$. Here recall that this term is indeed excluding  blocker job $b$.
Conditioned on the given state \emph{and} the blocker job turning out long, define $\Pi'$ to be SEPT from that point onwards, and denote by  $c_{\Pi'}(\text{remain})$ the corresponding expected remaining cost of $\Pi'$.
Note that in this case, $\Pi'$ has one machine \emph{less} available for the remaining jobs when compared to $\Pi$. Then the expected remaining cost for $\Pi'$ is at most 
\[
    (1-q)\cdot c_\Pi(\text{remain})\ +\ q\cdot c_{\Pi'}(\text{remain}).
\]
Here, the first term accounts for the event where blocker job $b$ turns out short, and 
the second term accounts for the event when blocker job $b$ turns out long.

Now it follows that  the expected cost of $\Pi$ minus the expected costs of $\Pi'$ is at least
\begin{equation}\label{eq:blockers_first}
    1/n+q\left(c_\Pi(\text{remain})-c_{\Pi'}(\text{remain})\right).
\end{equation}
Observe that $c_\Pi(\text{remain})-c_{\Pi'}(\text{remain})$ will generally be negative. However, if we choose $q$, the probability \red{for having a long blocker job}, small enough, then \eqref{eq:blockers_first} is strictly positive. Here,  $q=1/(\red{4}n^2B)$ is sufficient, because $c_\Pi(\text{remain})\ge 0$ and  $c_{\Pi'}(\text{remain})\le \red{4}nB$. \red{The latter bound holds because we have $2m=2n$ jobs in total, of maximal length $B+1+\frac{1}{n}$, hence each job has a start time  $\le B+1+\frac{1}{n}\le 2B$}. Hence, for small enough $q$, $\Pi'$ has strictly smaller expected cost in comparison to $\Pi$, contradicting $\Pi$ being optimal.
\end{proof}

With Lemma~\ref{lem:OPT_equal_ALG} at hand we can prove Theorem~\ref{thm:OPT}
\begin{proof}[Proof of Theorem~\ref{thm:OPT}]
    We argue by considering the same two instances that were used in order to prove Theorem~\ref{thm:SEPT}. By choosing a small enough probability $q$ for the blocker jobs being long,  Lemma~\ref{lem:OPT_equal_ALG} shows that 
    SEPT is an optimal scheduling policy.
    This is true because the knapsack jobs together with the dummy job are stochastically comparable, and then for arbitrary machine availability times, SEPT is an optimal scheduling policy~\cite{WVW86}.
    As a consequence, we can replicate the same arguments as in Theorem~\ref{thm:SEPT}: 
    The only case when there is a nonzero contribution to the difference $\Delta$, is when \emph{all} blocker jobs are long. In that case, again the optimum policy schedules the knapsack jobs identically in SEPT order on the free machine in both instances, as long as that machine still offers a starting time $\le B+1$. 
    Hence exactly like in the proof of Theorem~\ref{thm:SEPT}, for each NO realization, there is precisely one job that will start $1/n$ time earlier in Instance~1, than it does in Instance~2. 
    This allows us to do the counting of NO instances for  the given knapsack constraint. So assuming we could compute the expected cost that is achieved by an optimal scheduling policy, we could compute the number of feasible solutions for {\sc Knapsack}. To see that all necessary computations are indeed polynomial time, it suffices to observe that with the chosen probability $q=1/(4n^2B)$, we have that {the encoding length of probability $q^{m-1}$ is}  $|q^{m-1}|\in\bigO{m(\log n + \log B)}$.
\end{proof}

Using Theorem~\ref{thm:OPT}, we can finally derive  our main result by binary search. 
\begin{theorem}\label{thm:OPT2}
    For stochastic parallel machine scheduling to minimize $\E{\sum_j C_j}$,  the problem to decide if there exists a policy with expected cost $\E{\sum_j C_j}\le x$ for any given instance and $x>0$, is a \nump-hard problem.
\end{theorem}
\begin{proof}
    The idea is to make use of Instance~1 and Instance~2 in the proof of Theorem~\ref{thm:OPT}, and argue that an oracle to decide if a policy exists with $\E{\sum_j C_j}\le x$, \red{or equivalently 
    $\E{\sum_j S_j}\le x-\E{\sum_j X_j}$}
    for any $x>0$, allows to compute the exact value of the minimum expected cost for both instances 
    by  binary search. Theorem~\ref{thm:OPT} says that this is a \nump\-hard problem. 
    The reason why this works is the following: Both instances have $m(=n)$ machines, $m+n$ jobs, and the following properties:
    \begin{itemize}
        \item Every realized job size is a multiple of $\frac{1}{n}$ in the range $[0, B + 1 + \frac{1}{n}]$, 
        \item the probability of any joint realization of all job sizes is a multiple of $q^{m-1} \cdot (\frac{1}{2})^n$, where we take $q = \frac{1}{\red{4}n^2 B}$,\footnote{\red{Indeed, since $q=1/k$ for $k\in\mathbb{Z}_{\ge 0}$, $(1-q)=(k-1)/k$ is a multiple of $q$. Therefore, for $0\le h\le m-1$, the probability for a specific realization with $h$ short blocker jobs equals $q^{m-h-1}(1-q)^{h}= (k-1)^h(1/k)^{m-1}$, a multiple of $(1/k)^{m-1}=q^{m-1}$.}}
        \item \red{the total expected cost $\E{\sum_j S_j}$ is trivially upper bounded by $4nB$, because none of the $2n$ jobs starts later than $2B$}
    \end{itemize}
    Now by Lemma~\ref{lem:OPT_equal_ALG} and the proof of Theorem~\ref{thm:OPT} we know that we can restrict to non-idling policies for these two instances. That implies that  we have a finite and discrete set of possible values for the expected cost $\E{\sum_j S_j}$, and the expected cost of any non-idling and non-optimal policy exceeds that of an optimal policy by at least 
\[
q^{m-1}\cdot\left(\frac12\right)^n\cdot\frac1n\,.
\]
    Here, the product of the first two terms is a lower bound on the probability of any event, and the third term is the smallest realized processing time. 
    That implies that the number of required iterations of binary search to find the exact optimum value for these two instances is logarithmic in the above term, which is still polynomial in $n,m$, and $\log B$. Now Theorem~\ref{thm:OPT} yields the claimed result.
\end{proof}

\section*{Conclusions}
The hardness results of Theorems~\ref{thm:OPT} and~\ref{thm:OPT2} still allow that there is an algorithm that computes an optimal scheduling policy $\Pi$ in polynomial time. Here, one may imagine $\Pi$ as some polynomial space automaton that can decide, for any given state, which job should be scheduled next. In fact, (W)SEPT, or any other list scheduling policy, can be thought of as a simple version of such automaton, and indeed, we can easily compute the (W)SEPT order, yet this paper shows that we cannot, in general, compute its expected cost. We leave it as an interesting open question to show the hardness of computing 
the optimal policy $\Pi$ itself.

\subsection*{Acknowledgements} Marc Uetz acknowledges the hospitality at CMU and UPitt during the summer 2025, and the financial support by the EEMCS faculty of University of Twente. We thank Bodo Manthey for a helpful discussion on the proof of Lemma~\ref{lem:pack}.  We also acknowledge some helpful reviewer comments.

\bibliographystyle{plain}
\bibliography{SEPT}

\appendix

\section{Deferred Proofs}
\subsection{Proof of Fact~\ref{fact:minimal_idleness}}
\label{app:fact1}
The proof of Fact~\ref{fact:minimal_idleness} follows by a simple exchange argument.
\begin{proof}
 Consider a policy $\Pi$ that, at some time $t\ge 0$, leaves an idle time of length $\delta$, which is less than the time until the next potential job completion. And assume w.l.o.g.\ it schedules some job after that idle time on the same machine. Let $k$ be the first such job. Then define another policy that schedules job $k$ directly at time $t$, and the same idle time of length $\delta$ directly after job $k$ is finished. This policy still non-anticipatory, because there is no information gain between $t$ and $t+\delta$. Moreover, it achieves strictly better objective value than $\Pi$.
\end{proof}

\subsection{Proof of Lemma~\ref{lem:pack}}
\label{app:Lemma1}
The \nump-hardness of the restricted version of {\sc Knapsack} follows by reduction from {\sc Knapsack}. 

\begin{proof}
    First, it is well known that  {\sc \#Knapsack}, so the problem to count the number of feasible solutions for a knapsack constraint, is \nump-hard. Let us briefly recall why: It follows from the NP-hardness of the problem to decide if there exists a solution $W$ with $\sum_{i\in W}s_i =B$~\cite{Karp1972}, and because the corresponding NP-hardness reduction is parsimonious~\cite{Simon77}. This yields \nump-hardness of the problem to compute the number of $W$ with $\sum_{i\in W}s_i =B$.
    Consequently, counting the solutions $W$ with $\sum_{i\in W}s_i \le B$ is \nump-hard, too: Solving the counting problem for $\le$ thresholds $B$ and $B-1$, and subtracting, gives the number of $W$ with  $\sum_{i\in W}s_i \red{=} B$.
    As to the second part of Lemma~\ref{lem:pack}, first note that when $\sum_{i\in[n]}s_i\le B+1$, by integrality there are precisely $2^n-1$ or $2^n$ feasible solutions, hence w.l.o.g.\ we may assume $\sum_{i\in[n]}s_i> B+1$.
    We aim to show that also the constraint $\sum_{j\in[n]} s_j \le 3B/2$ is w.l.o.g. To that end, observe that any instance $I=(B, s_1,\dots,s_n)$  with $S:=\sum_i s_i$ and $f$ feasible knapsack solutions,  
    can be transformed into an equivalent instance $I'$ where  $B'$ is only an epsilon fraction smaller than $S'$: Add to the $n$ given items a set of $k$ dummy items with sizes $s'_i=S$, so that $S'=(k+1)S$ and let $B':=kS+B$. Then  $B'\ge \frac{k}{k+1} S'$, and the number of feasible solutions for $I'$ is $f'=(2^k-1)2^n+f$, because as soon as at least one dummy item is \emph{not} included, all other items fit.
    Hence the counting problem for instance $I'$ also solves the counting problem for the original instance $I$. Letting $k=2$ gives precisely $\sum_{j\in[n]} s_j \le 3B/2$.
\end{proof}

\end{document}